\documentclass[fleqn,usenatbib]{mnras}

\usepackage{newtxtext,newtxmath}

\usepackage[T1]{fontenc}
\usepackage{ae,aecompl}
\usepackage{siunitx}
\usepackage[para]{threeparttable}

\usepackage{graphicx}	
\usepackage{amsmath}	
\usepackage{wrapfig,lipsum}
\usepackage{hyperref}
\usepackage{subfigure}
\usepackage{upgreek}

\title[MWA observation of GRB 210419A]{High time resolution search for prompt radio emission from the long GRB 210419A with the Murchison Widefield Array}

\author[J. Tian et al.]{J. Tian$^{1}$\thanks{E-mail: jun.tian@postgrad.curtin.edu.au}, G. E. Anderson$^1$, P. J. Hancock$^1$, J. C. A. Miller-Jones$^1$, M. Sokolowski$^1$, 
\newauthor N. A. Swainston$^1$, A. Rowlinson$^{2,3}$, A. Williams$^1$, D. L. Kaplan$^4$, N. Hurley-Walker$^1$, 
\newauthor J. Morgan$^1$, N. D. R. Bhat$^1$, D. Ung$^1$, S. Tingay$^1$, K. W. Bannister$^5$, M. E. Bell$^6$, \newauthor B. W. Meyers$^7$, M. Walker$^1$\\
$^1$International Centre for Radio Astronomy Research, Curtin University, GPO Box U1987, Perth, WA 6845, Australia\\
$^2$Anton Pannekoek Institute, University of Amsterdam, Postbus 94249, 1090 GE, Amsterdam, The Netherlands\\
$^3$ASTRON, the Netherlands Institute for Radio Astronomy, Oude Hoogeveensedijk 4, 7991 PD, Dwingeloo, The Netherlands\\
$^4$Department of Physics, University of Wisconsin-Milwaukee, 1900 E. Kenwood Boulevard, Milwaukee, WI 53211, USA\\
$^5$Australia Telescope National Facility, CSIRO Astronomy and Space Science, PO Box 76, Epping, NSW 1710, Australia\\
$^6$University of Technology Sydney, 15 Broadway, Ultimo NSW 2007, Australia\\
$^7$Department of Physics and Astronomy, University of British Columbia, 6224 Agricultural Road, Vancouver, BC V6T 1Z1, Canada
}%

\date{Accepted XXX. Received YYY; in original form ZZZ}

\pubyear{2021}

\begin{document}
\label{firstpage}
\pagerange{\pageref{firstpage}--\pageref{lastpage}}
\maketitle

\begin{abstract}
We present a low-frequency (170\textendash200\,MHz) search for prompt radio emission associated with the long GRB 210419A using the rapid-response mode of the Murchison Widefield Array (MWA), triggering observations with the Voltage Capture System (VCS) for the first time. The MWA began observing GRB 210419A within 89\,s of its detection by \textit{Swift}, enabling us to capture any dispersion delayed signal emitted by this GRB for a typical range of redshifts. We conducted a standard single pulse search with a temporal and spectral resolution of $100\,\upmu$s and 10\,kHz over a broad range of dispersion measures from 1 to $5000\,\text{pc}\,\text{cm}^{-3}$, but none were detected. However, fluence upper limits of $77\text{--}224$\,Jy\,ms derived over a pulse width of $0.5\text{--}10$\,ms and a redshift of $0.6<z<4$ are some of the most stringent at low radio frequencies. 
We compared these fluence limits to the GRB jet-interstellar medium (ISM) interaction model, placing constraints on the fraction of magnetic energy ($\epsilon_{\text{B}}\lesssim[0.05 \text{--} 0.1]$). We also searched for signals during the X-ray flaring activity of GRB 210419A on minute timescales in the image domain and found no emission, resulting in an intensity upper limit of $0.57\,\text{Jy}\,\text{beam}^{-1}$, corresponding to a constraint of $\epsilon_{\text{B}}\lesssim10^{-3}$.
Our non-detection could imply that GRB 210419A was at a high redshift, there was not enough magnetic energy for low-frequency emission, or that the radio waves did not escape from the GRB environment.
\end{abstract}




\section{INTRODUCTION }

Since the discovery of gamma-ray bursts \citep[GRBs;][]{Klebesadel73, Mazets74}, our understanding of these events and their progenitors has been steadily increasing. Two categories of GRBs have been identified in the large population of GRBs: long and short, with pulse durations longer or shorter than 2\,sec, respectively \citep{Kouveliotou93}. Their progenitors are also different. While short GRBs have been confirmed to originate from compact binary mergers by the near-coincident detection of GRB 170817A \citep{Goldstein17} and GW170817 \citep{Abbott17}, long GRBs are commonly associated with core-collapse supernovae (e.g. \citealt{Hjorth03}; \citealt{Cano13}).

Radio synchrotron emission has been observed during the $\sim$1\textendash1000\,day afterglow phase of long GRBs \citep{Chandra12, Anderson18} and $\sim$1\textendash10\,day afterglow phase of short GRBs \citep{Fong15, Fong21, Anderson21b}. This emission likely results from relativistic jet ejecta interacting with the circum-burst media \citep{Sari95, Meszaros97}. There may also be two distinct populations of GRBs: radio bright and radio faint, which differ in their gamma-ray fluences, isotropic energies and X-ray fluxes, suggesting different prompt emission mechanisms or central engines \citep{Hancock13}.

In the standard fireball model \citep{Cavallo78, Rees92}, GRBs are supposed to launch relativistic jets by collapsars or binary mergers. Whether the GRB jets are Poynting flux or baryon dominated is still under debate \citep{Sironi15}. If the GRB jet is Poynting flux dominated, the magnetic energy is much larger than the particle energy \citep{Usov94, Thompson94, Drenkhahn02}, and its interaction with the ISM is predicted to generate a low frequency, coherent radio pulse at the highly magnetised shock front through magnetic reconnection, the same emission mechanism as for the gamma-ray emission \citep{Usov00}.

The central engine of long GRBs could be either magnetars or black holes formed via the core collapse of massive stars (for a review see \citealt{Levan16}). 
The formation of magnetar remnants by long GRBs is supported by X-ray observations. Of the X-ray afterglows detected by the \textit{Neil Gehrels Swift Observatory} (hereafter referred to as \textit{Swift}; \citealt{Gehrels04}) from many GRBs, a considerable fraction show a plateau phase during the X-ray decay \citep{Evans09}, suggesting continued energy injection after the prompt emission phase of the GRB \citep{Johannesson06}. Both the core collapse (long) and the binary merger (short) may form a quasi-stable, highly magnetised, rapidly rotating neutron star (magnetar), which could supply the necessary energy to power the plateau phase (e.g. \citealt{Troja07}; \citealt{Lyons10}; \citealt{Rowlinson13}).

In the case of a magnetar remnant, the coherent radio emission powered by dipole magnetic braking may appear as persistent or pulsed emission during the lifetime of the magnetar \citep{Totani13}. Depending on the equation of state of nuclear matter \citep{Lasky14}, the magnetar remnant may eventually spin down to a point at which it cannot be centrifugally supported, and thus collapses into a black hole, producing a final prompt radio signal due to magnetic field shedding \citep{Zhang14}. The prompt radio signals predicted for long GRBs may not escape their dense circum-burst environments if this coherent radiation is emitted below the plasma frequency \citep{Zhang14}. Nonetheless, it is possible that instances of lower circum-burst densities and different viewing angles may allow the signal to escape.
Additionally, since the effective optical depth for a single short pulse is determined by the duration of the pulse rather than the scattering medium, a short enough pulse could propagate out from the central engine \citep{Lyubarsky08}.

In the GRB X-ray light curves, there is another important feature that could help understand the central engines, X-ray flares, which are erratic temporal features superimposed on the regular decay (e.g. \citealt{Campana06}; \citealt{Falcone06}; \citealt{Margutti11}). They usually occur from $10^2$ to $10^5$\,s after the prompt emission, with a fluence usually lower than that of the prompt gamma-ray emission and a temporal behavior similar to the gamma-ray pulses (e.g. \citealt{Chincarini07, Chincarini10, Falcone07}). Thanks to the short slew time of the \textit{Swift} X-ray Telescope (XRT; \citealt{Burrows05}), X-ray flares following the prompt gamma-ray emission are commonly observed among \textit{Swift}\textendash BAT triggered GRBs (48\%; \citealt{Swenson14}). 
The same mechanism predicted to produce prompt radio emission when a Poynting flux dominated gamma-ray jet impacts the ISM  \citep{Usov00} may also apply to these X-ray flares if they are generated by the same internal shock mechanism \citep{Starling20}.

The prompt, coherent signals predicted to be produced by GRBs may be similar to fast radio bursts (FRBs) with millisecond durations \citep{Chu16, Rowlinson19}. There are potentially two classes of FRBs: repeaters and non-repeaters (e.g. \citealt{CHIME21}). Their origin is still unclear and there could be more than one source/progenitor population (for a review see \citealt{Zhang20c}). Currently there is compelling observational evidence that links both repeating and non-repeating FRBs to magnetar engines \citep{CHIME20, Bochenek20, CHIME21b}. Given GRBs could be the progenitors of magnetars, it is natural to make a connection between GRBs and FRBs (e.g. \citealt{gourdji20}). Therefore, detections of FRB-like emission associated with GRBs would support that such events can produce magnetar remnants, and indicate whether these signals may make up a subset of the FRB population.

Most of the FRB detections have been made at frequencies above 400\,MHz (the bottom of the Canadian Hydrogen Intensity Mapping Experiment/FRB observing band; \citealt{CHIME19}). However, the most recent detections by the LOw Frequency ARray (LOFAR; \citealt{Haarlem13}) at 110\textendash188\,MHz \citep{Pleunis21} of the repeating FRB 20180916B are by far the lowest frequency detections of any FRB to date, confirming the existence and the detectability of FRB-like emission from cosmological distances (not limited by propagation effects or the FRB emission mechanism - at least in the case of repeating FRBs) at low frequencies.

There have been searches for prompt radio emission from GRBs, which could be associated with the central engine or the relativistic jet, but so far none have yielded a detection (e.g. \citealt{Dessenne96}; \citealt{Obenberger14}; \citealt{Bannister12}; \citealt{Palaniswamy14}; \citealt{Kaplan15}; \citealt{Anderson18}; \citealt{Rowlinson19b}; \citealt{Anderson20}; \citealt{Bouwhuis20}; \citealt{Rowlinson21}; \citealt{Tian22}). These non-detections could be attributed to the small sample size (not all GRBs 
are predicted to 
produce detectable radio emission), the limited sensitivity of 
all-sky instruments, or long 
delays (up to several minutes) between the transient event and the target acquisition of pointed telescopes. 
Of these previous searches, very few were conducted at a temporal resolution sufficient for resolving prompt signals ($\sim$\,ms). \citet{Bannister12} performed a search at 1.4\,GHz with a time resolution of $64\,\upmu$s to 1\,s 
for prompt emission from 
nine GRBs (seven long and two short) and found two possible dispersed radio pulses associated with the X-ray plateau phases of two long GRBs, however, they are unlikely to be real due to their low significance and could be radio frequency interference (RFI) contamination, resulting in a fluence limit of $10\text{--}227$\,Jy\,ms depending on the pulse width. \citet{Palaniswamy14} performed another search at 2.3\,GHz with a time resolution of $640\,\upmu$s to 25.6\,ms 
for prompt emission from 
five long GRBs but did not detect any events above $6\,\sigma$, corresponding to a fluence limit of $75$\,Jy\,ms on 25\,ms timescales.

It is noteworthy that the LOFAR has been used to trigger rapid-response observations on \textit{Swift} GRBs, yielding the deepest upper limits to date for associated coherent, persistent radio emission from a magnetar remnant. Its observations of GRB 180706A (long) and 181123B (short) presented flux density limits of 1.7\,mJy and 153\,mJy, respectively, over a 2\,hr timescale~\citep{Rowlinson19b, Rowlinson21}.

The Murchison Widefield Array (MWA; \citealt{Tingay13}, \citealt{Wayth18}) has been performing triggered observations of both long and short GRBs since 2015 using the standard correlator (imaging mode), which has a temporal resolution of 0.5\,s (e.g. \citealt{Kaplan15}). In 2018, the MWA triggering system was upgraded to enable it to trigger on Virtual Observatory Events (VOEvents; \citealt{Seaman11}), allowing the MWA to point to a GRB position and begin observations within 20\,s of receiving an alert \citep{Hancock19}. This rapid-response system, triggering on both \textit{Swift} and \textit{Fermi} GRBs, makes the MWA a competitive telescope in searching for the earliest radio signatures from GRBs. As radio signals are expected to be delayed by the intervening medium between their origin and the Earth, with lower frequency signals arriving later, the low operational frequency of MWA (80--300\,MHz; \citealt{Tingay13, Wayth18}) makes it possible to catch a signal emitted simultaneously to, or even before, a GRB. 

The first triggered MWA observation on a short GRB was performed by \citet{Kaplan15}, and yielded an upper limit of 3\,Jy on 4\,s timescales. \citet{Anderson20} reported the first short GRB trigger with the upgraded MWA triggering system, and performed a search for dispersed signals using images with a temporal and spectral resolution of 0.5\,s/1.28\,MHz, obtaining a fluence upper-limit range from 570\,Jy\,ms at a dispersion measure (DM) of $3000$\,pc\,cm$^{-3}$ ($z\sim 2.5$) to 1750\,Jy\,ms at a DM of $200$\,pc\,cm$^{-3}$ ($z\sim 0.1)$, corresponding to the known redshift range of short GRBs \citep{Rowlinson13}. Finally, \citet{Tian22} presented a similar search in the image domain for coherent radio emission from a sample of nine short GRBs between 2017 and 2020 using the MWA rapid-response observing mode and obtained the most stringent prompt emission fluence limit 
of 100\,Jy\,ms from GRB 190627A.

In 2021, the MWA completed implementing a new triggering mode which incorporated the Voltage Capture System (VCS; \citealt{Tremblay15}). Unlike the standard correlator, the VCS records channelised data (3072 channels across 30.72\,MHz of bandwidth) for each tile ($4\times4$ dipole array) instead of correlated visibilities. 
This allows for the capture of high temporal and frequency resolution 
raw voltage data ($100\,\upmu$s/10\,kHz). Therefore, compared to previous works that use the MWA standard mode to trigger on GRBs with a temporal resolution of 0.5\,s \citep{Kaplan15, Anderson20, Tian22}, the triggered VCS observations are 
almost one order of magnitude more sensitive to millisecond duration transient signals (for a comparison of sensitivity between the standard MWA correlator and the VCS see figure 8 in \citealt{Tian22}). The new system triggers only on \textit{Swift} GRBs with $\sim$arcsec localisations. This enables us to localise the GRB to within a synthesised beam of the MWA and coherently beamform the VCS data, which can maximize our sensitivity. Given that the data rate of VCS observations is extremely large ($\sim$28\,TB/hr), we are not currently able to continuously observe one GRB for more than $\sim$100\,min \citep{Tremblay15}, making it difficult for us to detect prompt signals predicted to be produced at late times (e.g. by the collapse of an unstable magnetar remnant into a black hole, which may not occur for up to 2\,hr post-burst; \citealt{Zhang14}). However, VCS triggered observations of GRBs is the most promising method for searching for associated early-time prompt, coherent emission.

This paper presents the first GRB trigger with the MWA VCS mode and a search for prompt low frequency radio emission associated with GRB 210419A. In Section~\ref{sec:Obs}, we describe the observation of GRB 210419A obtained using \textit{Swift} and the VCS triggering mode of the MWA, and the data processing and analysis we used to search for prompt radio emission. Our results are then presented in Section~\ref{sec:results}. We use the upper limit derived from our VCS observation of GRB 210419A to constrain coherent radio signals associated with the relativistic jet during the prompt gamma-ray emission phase and from an X-ray flare 
in Section~\ref{sec:disc}.

\section{observations and analysis}\label{sec:Obs}
In this section, we describe our observation of GRB 210419A with the MWA rapid response mode, and introduce the data processing pipeline and the software we employed to perform a single pulse search for prompt radio emission associated with this GRB.

\subsection{\textit{Swift} observations}\label{sec:Swift}

\begin{figure*}
\centering
\includegraphics[width=.7\textwidth]{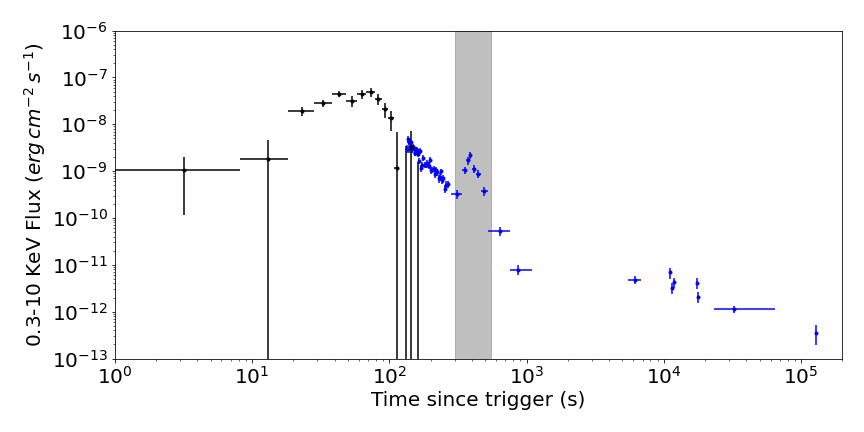}
\caption{0.3--10\,keV flux light curve of GRB 210419A. The black and blue data points were obtained by the \textit{Swift}-BAT (extrapolated to 0.3-10\,keV) and the \textit{Swift}-XRT, respectively. The shaded region indicates the period covering the X-ray flare investigated in Section~\ref{sec:model2}. The X-ray plateau phase starts around 1000\,s post-burst.} 
\label{X-ray}
\end{figure*}

The GRB 210419A was first detected by \textit{Swift} Burst Alert Telescope (BAT; \citealt{Barthelmy05}) at 06:53:41 UT on 2021 April 19 (trigger ID 1044032; \citealt{Laha21}). Refined analysis of the BAT light curve determined a $T_{90}$ of $64.43\pm11.69$\,s \citep{Palmer21}, unambiguously placing this GRB in the long GRB category ($T_{90}\gtrsim2$\,s; \citealt{Kouveliotou93}).
The time-averaged gamma-ray spectrum from T+21.92 to T+95.01\,s is best fit by a simple power-law model with an index of $2.17\pm0.24$, consistent with typical long GRBs \citep{Lien16}, and the gamma-ray fluence in the 15\textendash150\,keV band is $(7.8\pm1.2)\times10^{-7}\,\text{erg}\,\text{cm}^{-2}$ \citep{Palmer21}.

A subsequent detection of the X-ray afterglow by the \textit{Swift}-XRT localised this GRB to the position $\alpha(\text{J2000.0}) = 05^h47^m24^s.23$ and $\delta(\text{J2000.0}) =$ \ang[angle-symbol-over-decimal]{-65;30;09.0} with an uncertainty of \ang[angle-symbol-over-decimal]{;;2.0} (90\% confidence; \citealt{Osborne21}).
The XRT X-ray spectrum ($0.3-10$\,keV) at $\sim1$\,h post-burst is best fit by a power law with a photon index of $2.60^{+0.29}_{-0.27}$ and an absorption column of $1.9^{+0.7}_{-0.6}\times10^{21}\,\text{cm}^{-2}$ in the Photon Counting (PC) mode \citep{Beardmore21}.

\begin{figure*}
\centering
\includegraphics[width=.7\textwidth]{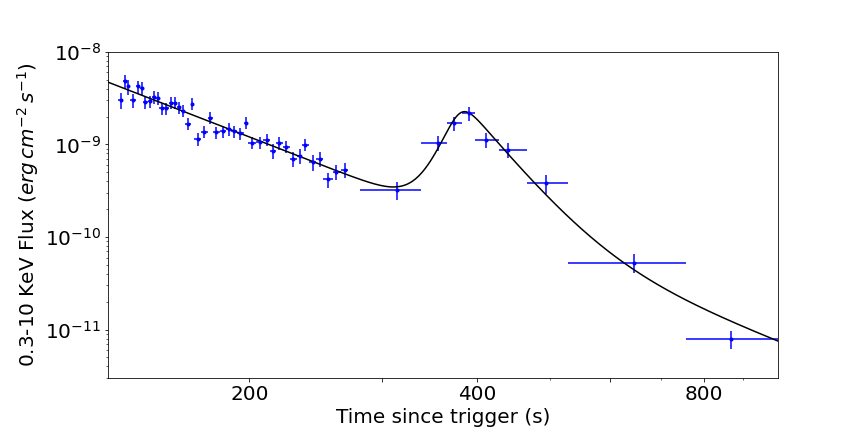}
\caption{The fit to the X-ray flare from GRB 210419A used to calculate its duration of 248\,s (between 335--583\,s post-burst). We used a smooth broken power-law function to fit the flare plus a declining power-law to fit the underlying X-ray light curve observed by \textit{Swift}\textendash XRT (blue points), with the fitting result being shown as the black line.} 
\label{Xfitting}
\end{figure*}

Combining the \textit{Swift}\textendash BAT and \textendash XRT data from the \textit{Swift} Burst Analyser \citep{Evans10}, we created the X-ray light curve for GRB 210419A in the 0.3\textendash10\,keV energy band in the observer frame, as shown in Figure~\ref{X-ray}. The light curve is characterised by a power law decay with an X-ray flare peaking at $\sim4\times10^2$\,s (shaded region), 
followed by a plateau phase starting from $\sim10^3$\,s. 
In order to calculate the duration and fluence of the X-ray flare of GRB 210419A, we fitted the flare with a smooth broken power-law function plus a declining power-law to model the underlying X-ray emission decay (see eq. 1 \& 2 in \citealt{Yi16}), as shown in Figure~\ref{Xfitting}. 
The duration of the flare (248\,s) is defined as the interval between the two intersections of the flare component and the underlying power law decay (335--583\,s; \citealt{Yi16}). Integrating the flux density over this duration, we obtained a fluence of $1.58\times10^{-7}\,\text{erg}\,\text{cm}^{-2}$ for the X-ray flare, which is a typical value among observed X-ray flares (see figure 1 in \citealt{Starling20}). For the analysis, results and interpretation of the X-ray flare see Sections~\ref{sec:imaging}, \ref{sec:results2} and \ref{sec:model2}, respectively.
We do not see a steep decay following the plateau phase until $\sim10^5$\,s, which might suggest it is powered by a stable magnetar \citep{Rowlinson13}.
No redshift was obtained for GRB 210419A. An optical follow-up of this GRB with the Las Cumbres Observatory (LCO) 1-m Sinistro instrument did not detect any uncatalogued optical source within the XRT error region \citep{Strausbaugh21}.

\subsection{MWA observations of GRB 210419A}\label{sec:obs}

The MWA triggered observation of GRB 210419A was taken at a central frequency of 185\,MHz with a bandwidth of 30.72\,MHz in the phase II compact configuration \citep{Wayth18} using the VCS mode, which has a temporal and frequency resolution of $100\,\upmu$s and 10\,kHz, respectively. The size of the MWA synthesised beam in this configuration is $\sim10$\,arcmin, much larger than the GRB positional error. The GRB position was continuously observed for 15\,min. Note that 45/128 tiles were offline 
during this observation, which resulted in 
a noticeable sensitivity loss (see Section~\ref{sec:sensi}).

The VOEvent broadcasting the \textit{Swift}\textendash BAT detection of GRB 210419A 
was circulated 76\,s post-burst.
Just 0.5\,s later,
the MWA rapid-response front-end web service received the VOEvent. The VOEvent handler took 0.3\,s to parse this VOEvent, identifying it as a real GRB, and triggered MWA VCS observations, with the
subsequent update of the 
MWA observing schedule taking 
1.3\,s.
The telescope then took a further 10.9\,s to re-point and begin collecting data in the VCS mode.
Therefore, the total latency between the \textit{Swift} broadcast of the VOEvent and the MWA being on target was 13\,s. Overall, the MWA observation of GRB 210419A started at 06:55:10 UT, just 89\,s following the \textit{Swift} detection. A time-line of the triggering process is summarised in Table \ref{trigger}.

\begin{table*}
\begin{threeparttable}
\centering
\resizebox{1.6\columnwidth}{!}{\hspace{+0cm}\begin{tabular}{c c c c}
\hline
 UT Time & Latency & Event & Description \\
 (2021-04-19)&(s) &(\#)  &  \\
\hline
06:53:41 & 0 & 1 & \textit{Swift}\textendash BAT detects GRB 210419A \\
06:54:57 & 76 & 2 & \textit{Swift} VOEvent alert notice circulated \\
06:54:57.5 & 76.5 & 3 & MWA front-end receives VOEvent \\
06:54:57.8 & 76.8 & 4 & VOEvent handler parses VOEvent and sends trigger to schedule observations \\
06:54:59.1 & 78.1 & 5 & MWA schedule is updated \\
06:55:06.2 & 85.2 & 6 & MWA is on target \\
06:55:10.0 & 89 & 7 & MWA sets up the VCS mode and begins observations \\
\hline
\end{tabular}}
\caption{Timeline for MWA triggered observation of GRB 210419A
}
\label{trigger}
\end{threeparttable}
\end{table*}

\subsection{Data processing}

In the following we describe the VCS data processing pipeline, including downloading, calibration, and beamforming (for details about the pipeline see \citealt{Bhat16}, \citealt{McSweeney17}, \citealt{Meyers17} and \citealt{Ord19}). Since GRB 210419A was localised to within the MWA phase II compact configuration tied-array beam, we could coherently beamform the data at the GRB position. The final data product is a time series of Stokes parameters packed into the \mbox{PSRFITS} format \citep{Hotan04}, which can be further analyzed by the {\sc presto} software package\footnote{\mbox{\url{https://github.com/scottransom/presto}}} \citep{Ransom01}. Here we present specific details regarding the calibration and beamforming of the VCS observation of GRB 210419A.

\subsubsection{Calibration}\label{sec:cal}
In order to coherently sum the voltages from the MWA tiles, we need to determine the direction independent complex gains, including amplitudes and phases, for each constituent tile 
(for details see \citealt{Ord19}). 
We selected a bright source (Hydra A) that had been observed in the standard correlator mode three hours after the GRB observation as the calibrator source. For each of the $24\times1.28$\,MHz sub-bands and each tile, a calibration solution for the amplitude and phase was generated from the visibilities using the Real Time System (RTS; \citealt{Mitchell08}). After inspecting each solution, we discarded a further nine tiles due to their poor calibration solutions. We also excised the edge channels (0\textendash7 and 120\textendash127) of each of the 24 sub-bands to alleviate the aliasing effects resulting from the channelisation process.

\subsubsection{Coherent beamforming}
The voltages from individual tiles can be coherently summed to form a tied-array beam (i.e., coherent beamforming). Compared to incoherent beamforming which simply sums up the power from each tile to preserve a large field of view, coherent beamforming can potentially gain more than an order of magnitude increase in sensitivity for each phase centered beam (a factor of $\sim10$ improvement in actual observations; \citealt{Bhat16}).
The performance of coherent beamforming is affected by a few factors, such as the quality of the calibration solution and the pointing direction of the telescope.

We used the coherent beamforming to phase all tiles to the GRB position. This requires the knowledge of cable and geometric delays to the pointing centre for each tile, which is then converted into phase shifts (for details see \citealt{Ord19}). Combining the delay model and the complex gain information from the calibration solution derived above, we obtained the tile based gain solution to phase all tiles to the same direction.

\subsubsection{Imaging over a long integration time}\label{sec:imaging}

While the high time resolution data are most suitable for searching for the prompt radio emission, snapshot images on $\sim$\,min timescales can be used for detecting dispersed long duration signals (see Section~\ref{sec:model2}). Compared to MWA imaging in the standard correlator mode (e.g., \citealt{Tian22}), there is a prerequisite step for imaging the VCS data, i.e. offline correlation for creating visibilities \citep{Susmita22}. We used the same calibration observation as in Section~\ref{sec:cal} for calibrating the visibilities, and made an image 
with 248\,s of data that covers the duration of the X-ray flare (highlighted in Figure~\ref{X-ray} and fitted in Figure~\ref{Xfitting}).
To take into account the dispersion delay in the arrival time of any associated radio emission with respect to the X-ray flare, we
offset the start time of the image.
Given the unknown redshift of GRB 210419A, we made images starting from 347\,s, 582\,s and 820\,s post-burst, corresponding to the dispersion delay for a typical long GRB at low, mean, and high redshift \citep{Le17}, including $z=0.1$, 1.7 and 4 respectively, and inspected them for any associated signals.


For the image we adopted a pixel scale of 2\,arcmin and 
size of $2048\times2048$ pixels, and used the {\sc WSClean} algorithm \citep{Offringa14, Offringa17} for deconvolution. 
This imaging exercise also provides a check on the data quality and the calibration solution despite the relatively poor imaging performance in the compact configuration.
The final MWA image covering the period of the X-ray flare assuming a redshift of $z=1.7$ 
is shown in Figure~\ref{image}. 
For the 
results and interpretation see
Sections~\ref{sec:results2} and \ref{sec:model2}.


\subsection{Data Analysis}

Using the VCS tied-array beamformer, we produced a time series with a temporal and frequency resolution of $100\,\upmu$s and 10\,kHz for our observation of GRB 210419A. 
As the prompt, coherent radio emission we are searching for will be dispersed in time by the medium it propagates through, it was necessary to perform a de-dispersion search to increase our sensitivity to any short-duration ($\sim$\,ms) signals. Considering this emission may be linked to different emission models and thus have different start times following the GRB (see Section~\ref{sec:model}), we performed the search across the entire 15\,min observation.

\subsubsection{Dispersed pulse search}\label{sec:search}

The de-dispersion search was performed using the {\sc presto} software package \citep{Ransom01}.
As the MWA is generally less affected by radio-frequency interference (RFI) when compared to other telescopes traditionally used for high-time resolution analysis, we did not perform any RFI removal that is 
often used at higher observing frequencies (see procedures outlined in \citealt{Swainston21}). Nevertheless, any spurious events caused by RFI can be identified from the final candidates by visual inspection.

We used the {\sc prepdata} routine in {\sc presto} to incoherently de-disperse the time series. Since there is no redshift measurement for GRB 210419A, we searched over a broad DM range from 1 to $5000\,\text{pc}\,\text{cm}^{-3}$, corresponding to a redshift range up to $z\sim4$ using the DM\textendash redshift relation $\text{DM}\sim1200z\,\text{pc}\,\text{cm}^{-3}$ (e.g. \citealt{Ioka03}; \citealt{Inoue04}; \citealt{Lorimer07}; \citealt{Karastergiou15}). This DM range covers up to 90\% of long GRBs detected by \textit{Swift} per year based on their known redshift distribution \citep{Le17}. The DM trials used for de-dispersion were determined by the {\sc ddplan.py} algorithm in {\sc presto} using the parameters passed: central frequency, bandwidth, number of channels, and sampling time. As the dispersive channel smearing increases with the DM value, the data were down sampled by up to a factor of 16 to match the smearing time. The DM step size was increased when the DM smearing would cause a loss in sensitivity equal to a DM error the size of half a DM step.
This results in 4401 DM trials and a temporal resolution ranging between 0.1\textendash1.6\,ms.

We searched for single pulses from each of the de-dispersed time series using {\sc presto}'s matched-filtering based {\sc single\_pulse\_search.py} routine, which convolves the time series with boxcars of different widths. To avoid missing any bright burst events, we disabled the check for bad blocks. Single pulse events detected with a signal-to-noise ratio (SNR) $>6$ were classified as candidates (e.g. \citealt{Chawla20}, \citealt{Meyers18}, \citealt{Bannister12}). We adopted the definition of $\sigma$, i.e. the noise level, as output by {\sc presto} for our SNR value (e.g. \citealt{Zhang20b}). Note that the statistics of this SNR may be complicated by a few facts, mainly that the matching algorithm is difficult to model statistically and the search over different DM trials and pulse widths is not necessarily independent (e.g. \citealt{Bannister12}).

Following this analysis, {\sc presto} identified 143 trials with a SNR$>6$, with a maximum SNR of 7.1 (for the distribution of the SNRs see Figure~\ref{hist} in Appendix~\ref{appendix:candi}). 
As a further test on the sample of $>6\sigma$ trials and the likelihood of some being real, 
we performed another single pulse search by creating a set of time series on the same dataset using (unphysical) negative DM trials \citep[see][for further details]{Tian22}. We found 119 candidates and a maximum SNR of 6.7 for the negative DM trials. Given the similar maximum SNR values resulting from the processing of both the positive and negative DM datasets, it is unlikely that any of the $>6\sigma$ candidates are real dispersed signals.

As a physically motivated filtering step, we examined the DM values of the candidates output by {\sc presto}.
Although it is difficult to predict the total DM of coherent emission associated with GRB 210419A as we do not know its redshift, we know that it is at cosmological distances. We can therefore use the DM contribution from the Milky Way in the GRB direction as a lower limit, which is $\text{DM}_{\text{MW}}\sim62\,\text{pc}\,\text{cm}^{-3}$ according to the YMW16 electron density model \citep{YMW16}.
All prompt signal candidates must therefore have a $\text{DM}>62\,\text{pc}\,\text{cm}^{-3}$.
We arrived at 11 candidates at this stage.

As a final filtering step, we used a friends-of-friends algorithm \citep{Burke11,Bannister12} to identify possible false positives. This algorithm exploits the fact that statistical fluctuations above the threshold are likely to appear only in a single DM trial and time stamp whereas a real signal would be partially detected in adjacent trials. Therefore, only candidates detected in a group of three or more adjacent DMs and times are likely to be real. Following both filtering criteria described above, there remained no valid candidate.

\subsubsection{Determination of system sensitivity}\label{sec:sensi}
We converted the $6\,\sigma$ threshold on the SNRs output by {\sc presto} into flux density limits using the radiometer equation,

\begin{equation}
    S_{\text{min}} = (\text{S}/\text{N})\times\frac{\text{SEFD}}{\sqrt{n_{\text{p}}t_{\text{int}}\Delta \nu}},
    \label{eq:sensi}
\end{equation}

\noindent where 
$n_{\text{p}}$ is the number of polarisations sampled, $t_{\text{int}}$ is the integration time in units of $\upmu$s, and $\Delta \nu$ is the bandwidth in units of MHz (see e.g. \citealt{Meyers17}).
The overall system equivalent flux density (SEFD) is determined by the ratio of the system temperature $T_{\text{sys}}$ and gain $G$, 
\begin{equation}
    \text{SEFD} = \frac{T_{\text{sys}}}{G} = \frac{\eta T_{\text{ant}}+(1-\eta)T_{\text{amb}}+T_{\text{rec}}}{G},
\end{equation}

\noindent where $\eta$ is the direction and frequency dependent radiation efficiency of the MWA array, and $T_{\text{ant}}$, $T_{\text{amb}}$ and $T_{\text{rec}}$ represent the antenna, ambient and receiver temperatures, respectively. The radiation efficiency $\eta$ at the position of GRB 210419A at our observing frequency of 185\,MHz is 0.987 \citep{Ung19}, the receiver temperature (which is well characterised across the MWA band) is 23\,K, and the ambient temperature (calculated from the metadata of our observation) is 311\,K.

The calculation of the antenna temperature and gain requires a good knowledge of the tied-array synthesised beam pattern, i.e. the product of the array factor and an individual MWA tile power pattern. The array factor contains the phase information that points the telescope to a target position, and the tile pattern can be simulated as described in \citet{Sutinjo15}. Assuming a sky temperature map at our observing frequency based on the global sky model of \citet{Costa08}, we convolved it with the tied-array beam pattern (e.g. \citealt{Sokolowski15}) to estimate the antenna temperature and the tied-array gain (for a full description of this procedure see \citealt{Meyers17}). Altogether, we found the SEFD for our coherently beamformed data to be 986\,Jy for the full (128 tiles) MWA.

We need to consider a few other factors in order to calculate our final sensitivity for this observation. First is the bandwidth consideration. As we flagged 16 of the 128 fine channels, the effective bandwidth is reduced to 87.5\% of the full 30.72\,MHz. To correct for this, we need to apply a scaling factor of $0.875^{-1/2}\approx1.07$ when converting to flux density limits. In estimating the SEFD, we used 128 tiles of the full MWA for our simulation. However, there were 45 bad tiles during our observation of GRB 210419A. In the ideal case where the sensitivity scales with the number of tiles, this means we have lost 35\% sensitivity.
Additionally, a coherency factor is introduced to quantify the deviation of the theoretical expectation from the actual improvement with respect to incoherent sums, and can be estimated by comparing the SNRs of a bright pulse in the coherently and incoherently beamformed data (for details see \citealt{Meyers17}). We chose the brightest pulsar (PSR J0437\textendash4715; \citealt{Bhat14}) in our field of view and produced an estimate of 0.639 for the coherency factor. This pulsar detection also demonstrates that our data processing and searching pipeline were operating correctly. Taking into account the above considerations, we arrived at a flux density upper limit of $6\sigma=25$\,Jy on a 1\,ms timescale.

To better characterise our sensitivity to prompt radio signals, we converted the flux density limit to a fluence limit

\begin{equation}
    F=25\,(w_{\text{obs}}/1\,\text{ms})^{1/2}\,\text{Jy\,ms},
    \label{eq:fluence}
\end{equation}

\noindent which is dependent on the pulse duration ($w_{\text{obs}}$). The observed pulse duration is given by

\begin{equation}
    w_{\text{obs}}=\sqrt{[w_{\text{int,rest}}(1+z)]^2+w_{\text{sample}}^2+w_{\text{DS}}^2+w_{\text{scatter}}^2},
    \label{eq:width}
\end{equation}

\noindent where $w_{\text{int,rest}}$, $w_{\text{sample}}$, $w_{\text{DS}}$ and $w_{\text{scatter}}$ are the rest-frame intrinsic pulse duration, observational sampling time, dispersion smearing, and pulse scattering, respectively \citep{Hashimoto20}. Here we assume that the scattering would not limit the observability of prompt radio signals \citep{Sokolowski18}. 
As the observed pulse width varies with the redshift, our prompt emission fluence limit is also redshift dependent (see Section \ref{sec:model1}).

\section{results}\label{sec:results}

\subsection{Prompt signal search}\label{sec:results1}

As described in Section~\ref{sec:search}, we performed a single pulse search on the high time resolution VCS triggered observation of \textit{Swift}-detected GRB 210419A. 
Of these 143 trials $>6\sigma$, only 11 had a DM$>62$\,pc\,cm$^{-3}$. As mentioned in Section~\ref{sec:search}, none of these candidates passed the friends-of-friends algorithm filtering, however, we still visually inspected the 11 candidates for signs of a dispersion sweep in the dynamic spectrum. None were seen  reaffirming they are not viable dispersed signal candidates.

In conclusion, for the DM range of 62--5000\,pc\,cm$^{-3}$, corresponding to all extragalactic distances up to $z\sim4$, we do not detect any associated prompt radio emission from GRB 210419A. For an intrinsic pulse width of $w_{\text{int,rest}}=0.5\text{--}10$\,ms (typical for FRBs; \citealt{Hashimoto19, Hashimoto20b}), our non-detection points to a $6\sigma$ fluence upper limit of 32--224\,Jy\,ms, which can be used to constrain theoretical coherent emission models (see Section~\ref{sec:model}).


\subsection{Long timescale emission during the X-ray flare}\label{sec:results2}

\begin{figure}
\centering
\includegraphics[width=.5\textwidth]{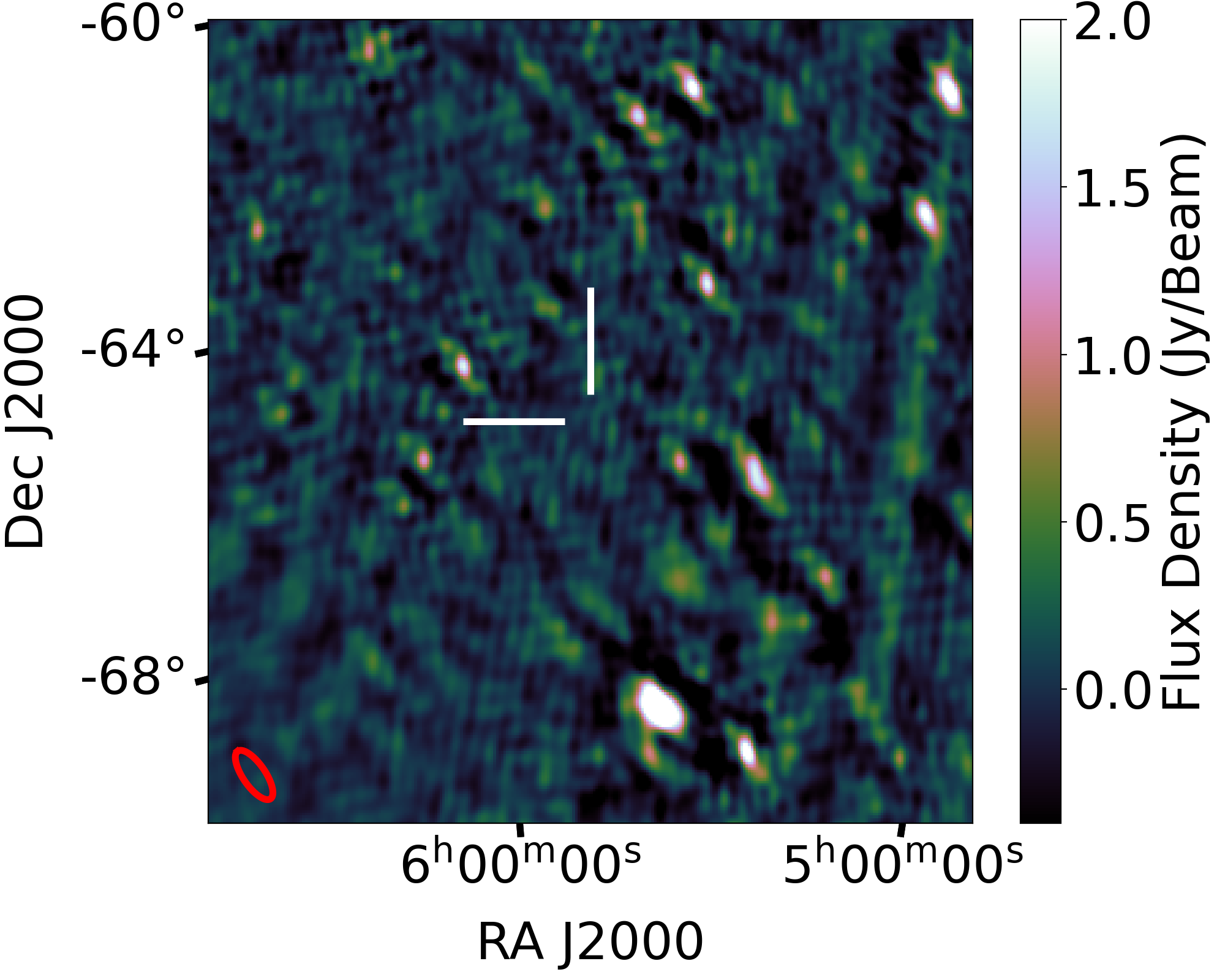}
\caption{The MWA image showing the region surrounding GRB 210419A integrated over the duration of the X-ray flare assuming a redshift of $z=1.7$ 
as described in 
Sections~\ref{sec:imaging} and \ref{sec:results2} (see radio fluence predictions in Section~\ref{sec:model2}). The two white lines point to the GRB position localised by the \textit{Swift}\textendash XRT to within a synthesised beam of the MWA, where the RMS noise was measured to be 190\,mJy\,$\text{beam}^{-1}$. The ellipse in the lower left corner shows the synthesized beam size of $23.4\times9.4$\,arcmin.} 
\label{image}
\end{figure}

In order to search for long duration coherent radio signals 
associated with
the 
X-ray flare 
described in Section~\ref{sec:Swift}, 
we generated an MWA image that covers the lifetime of the flare and potential dispersion delay for several redshifts 
as shown in Figure~\ref{image} (see Section~\ref{sec:imaging}). 
We performed a forced fit to the synthesised beam at the \textit{Swift}-XRT position of GRB 210419A using the radio transient detection work-flow {\sc Robbie}~\citep{Hancock19b}. 
For the three tested redshifts of $z=0.1$, 1.7, and 4, 
we obtained a flux density of 
$93\pm165\,\text{mJy}\,\text{beam}^{-1}$,
$115\pm169\,\text{mJy}\,\text{beam}^{-1}$,
 and $46\pm171\,\text{mJy}\,\text{beam}^{-1}$, respectively,
which are consistent with zero within the uncertainties and therefore indicating a non-detection. 
We used {\sc Robbie}~\citep{Hancock19b} to calculate 
a local RMS noise of $190\,\text{mJy}\,\text{beam}^{-1}$ in the image that assumes the GRB is at $z=1.7$ (Figure~\ref{image}), and therefore 
derived a $3\sigma$ upper limit of $570\,\text{mJy}\,\text{beam}^{-1}$ for the long timescale radio emission during the X-ray flare. The RMS noise in the other two images assuming redshifts of $z=0.1$ and $z=4$ are similar, $185\,\text{mJy}\,\text{beam}^{-1}$ and $192\,\text{mJy}\,\text{beam}^{-1}$, respectively.
The flux density upper limit 
can be used to constrain model parameters applicable to the GRB jet during the X-ray flare phase (see Section~\ref{sec:model2}).

\section{discussion}\label{sec:disc}

\subsection{Propagation effects}\label{sec:prop}

There are several propagation effects limiting the observability of the coherent, prompt radio emission we are searching for, such as 
absorption due to induced Compton scattering \citep{Condon16} and absorption below the plasma frequency 
in the dense environment of the emission site \citep{Condon16}. 
Such considerations are particularly import for long GRBs as observations show they often  
occur in star forming regions near the centres of their host galaxies (generally with low metallicity; e.g. \citealt{Berger09}, \citealt{Levesque10}), consistent with their core collapse origin. Due to the strong wind emission from a massive star prior to its collapse \citep{Weaver77}, the circum-burst media of long GRBs exhibit a large density range typically between $\sim10^{-1}\text{--}10^{2}\,\text{cm}^{-3}$ \citep{Laskar15}. 
In the following, we investigate the effect of both absorption mechanisms on any prompt radio emission emitted by GRB 210419A.

It has been shown that in the dense environments of long GRBs, induced Compton and Raman scattering can severely reduce the detectability of radio pulses at $\sim$MHz frequencies \citep{Macquart07}. For long GRBs in dense environments, only if the GRB jet is ultrarelativistic or the intrinsic opening angle of the emission is extremely small, could the predicted radio emission be visible. Given our incomplete knowledge of the GRB Lorentz factors (only lower limits have been observed, e.g. \citealt{Ackermann10}, \citealt{Zhao11} and \citealt{Zou11}) and the precise jet opening angles (they are likely confined to a narrow region; see \citealt{Beniamini19}; \citealt{Salafia20}), it is unknown whether the radio emission can evade induced Compton scattering. Nonetheless, a detection would provide valuable information on the Lorentz factor and the opening angle of the GRB jet. 
Specifically, 
it would indicate the evasion of induced Compton scattering,
implying that the intrinsic emission angle is less than or equal to 
$\Delta\Omega\lesssim5\times10^{-4}(T_\text{B}/10^{25}\,\text{K})^{-1/2}\,\text{sr}$ where $T_\text{B}=10^{24}\text{--}10^{29}$\,K is the brightness temperature of the radio emission \citep{Thompson94, Macquart07}. If the radio emission is isotropic in the rest frame of the jet, 
this means that the
minimum possible Lorentz factor of the jet is $\Gamma\gtrsim10^3(D/100\,\text{Mpc})$ where $D$ is the luminosity distance of the GRB \citep{Macquart07}.

The column density obtained from the X-ray spectrum of GRB 210419A can be used to estimate the plasma absorption of the radio emission along the line of sight.
The X-ray spectrum of GRB 210419A is best fitted with an absorbed power law with a photon index of $2.60^{+0.29}_{-0.27}$ and an absorption column of $1.9^{+0.7}_{-0.6}\times10^{21}\,\text{cm}^{-2}$ \citep{Beardmore21}, which is in excess of the Galactic value of $7.8\times10^{20}\,\text{cm}^{-2}$ \citep{Willingale13}. If we ignore the contribution of intervening systems (the interstellar and intergalactic media) along the line of sight, the intrinsic absorbing column density for GRB 210419A would be $\sim1.1\times10^{21}\,\text{cm}^{-2}$, which is smaller than the typical value of $5_{-3.4}^{+10.8}\times10^{21}\,\text{cm}^{-2}$ evaluated from a sample of long GRBs \citep{Campana12} and comparable to the typical value of $2.5_{-1.5}^{+3.8}\times10^{21}\,\text{cm}^{-2}$ for a sample of short GRBs \citep{Asquini19}.
As shown by \citet{Zhang14} in the specific context of GRBs, the plasma frequency in the GRB environment must be lower than the radio frequency for the radio emission to escape, i.e. $1/2\pi\times\sqrt{4\pi n_e e^2/m_e}<\nu_\text{obs}$, where $n_e$ is the electron number density, and $e$ and $m_e$ are the electric charge and mass of electrons \citep{Vlasov68}. At the MWA observing frequency of $\nu_\text{obs}=185$\,MHz, that would require an electron number density $n_e\lesssim4\times10^8\,\text{cm}^{-3}$, corresponding to an electron column density of $\lesssim4\times10^{21}\,\text{cm}^{-2}$ if we assume the length scale of the GRB environment to be $\sim10^{13}$\,cm \citep{Zhang14}. 
While the electron column density along our line of sight derived from the XRT spectrum of GRB 210419A
is less than this value, the uncertainty associated with the length scale makes it difficult to conclude whether our observing frequency is above the plasma frequency.
For the following analysis, we assume that it is above the plasma frequency in order to investigate the constraints our observations place on coherent radio emission predicted by the jet-ISM interaction model. 

\subsection{Constraints on the jet-ISM interaction model}\label{sec:model}

\subsubsection{Radio emission associated with the prompt gamma-ray emission}\label{sec:model1}
As suggested by \citet{Usov00}, the interaction between a Poynting flux dominated jet and the ISM can generate a coherent radio pulse as well as the prompt gamma-ray emission.
In this scenario, the bolometric radio fluence $\Phi_{r}$ ($\text{erg}\,\text{cm}^{-2}$) is proportional to the bolometric gamma-ray fluence $\Phi_{\gamma}$ ($\text{erg}\,\text{cm}^{-2}$) in the energy range of $0.1\text{--}10^4$\,keV, the widest energy range for current GRB detection satellites (e.g. \citealt{Rowlinson19b}). This power ratio is roughly estimated to be $\simeq0.1\epsilon_{B}$ \citep{Usov00}, where $\epsilon_{B}$ is the fraction of magnetic energy in the relativistic jet. In the typical spectrum of low-frequency waves generated at the shock front, there is a peak frequency determined by the magnetic field

\begin{equation}
    \nu_\text{max}\simeq[0.5\,\text{--}\,1]\frac{1}{1+z}\epsilon_{B}^{1/2}\times10^6\,\text{Hz}
    \label{eq:fpeak}
\end{equation}

\noindent (in the observer's frame; \citealt{Rowlinson19b}). For our observing frequency $\nu=185$\,MHz, which is above the peak radio frequency, the observed radio fluence is given by

\begin{equation}
    \Phi_{\nu}=\frac{\beta-1}{\nu_{\text{max}}}\Phi_{r}\bigg(\frac{\nu}{\nu_{\text{max}}}\bigg)^{-\beta}\,\text{erg}\,\text{cm}^{-2}\,\text{Hz}^{-1}.
\end{equation}

\noindent Note that the bolometric radio fluence $\Phi_r$ is the fluence integrated over frequency and thus has a different unit to $\Phi_\nu$. Assuming a typical spectral index of $\beta=1.6$ \citep{Usov00}, the power ratio between $\Phi_r$ and $\Phi_{\gamma}$ can be written in terms of the radio fluence at our observing frequency:

\begin{equation}
    \delta=\frac{5}{3}\nu^{1.6}\nu^{-0.6}_{\text{max}}\frac{\Phi_{\nu}}{\Phi_{\gamma}}.
\end{equation}

\noindent Thus, the predicted radio fluence is given by

\begin{equation}
    \Phi_{185\text{MHz}}\simeq[0.9\,\text{--}\,1.4]\times10^{-10}\delta(1+z)^{-0.6}\epsilon_B^{0.3}\Phi_{\gamma}.
    \label{eq:model}
\end{equation}

In order to calculate the unabsorbed bolometric gamma-ray fluence ($0.1-10^{4}$\,keV) for GRB 210419A, we applied a correction factor to the gamma-ray fluence measured by \textit{Swift}-BAT in the $15-150$\,keV energy band.
Assuming a simple power-law model for the spectrum of the prompt emission as given by \citet[][]{Palmer21}
and the absorption column derived from the spectral fit to the \textit{Swift}-XRT PC observation of GRB 210419A \citep[][see also Section~\ref{sec:Swift}]{Beardmore21}, we used the WebPIMMS tool \citep{Mukai93} to obtain a fluence correction factor of $6.2^{+6.5}_{-1.2}$. Note that the errors come from the uncertainty on the spectral index, which dominates the errors on the absorption column.  


Both the model-predicted prompt radio emission and our fluence upper limit at 185\,MHz (see Eq. \ref{eq:fluence}) depend on redshift, which is an unknown quantity for GRB 210419A. Under the assumption that we would be able to capture the dispersion delayed radio emission generated at the prompt gamma-ray emission phase (when the GRB jet first interacts with the ISM), the 89\,s delay of our observation with respect to the GRB detection (see Section \ref{sec:obs}) means we can only detect signals with a minimum DM of $734\,\text{pc}\,\text{cm}^{-3}$. After subtracting the Galactic contribution (see Section \ref{sec:search}), this corresponds to events at $z\gtrsim0.6$.
We are therefore able to search for prompt radio signals associated with the jet-ISM interaction within the redshift range of $0.6<z<4$ for GRB 210419A.

In order to constrain the model-predicted prompt emission in Eq. \ref{eq:model}, we need to convert the sensitivity of our observation to a fluence upper limit using Eq. \ref{eq:fluence}, which is dependent on 
the unknown rest-frame intrinsic pulse width $w_{\text{int,rest}}$. In the absence of detected prompt emission from long GRBs, we base our choice of $w_{\text{int,rest}}$ on known rest-frame intrinsic durations of FRBs with known redshifts and no scattering features ($\sim0.5\text{--}10$\,ms; \citealt{Hashimoto19, Hashimoto20b}).
We therefore assume durations of $w_{\text{int,rest}}=0.5$\,ms and 10\,ms for our fluence upper limits when constraining the model predictions.

\begin{figure}
\centering
\includegraphics[width=.5\textwidth]{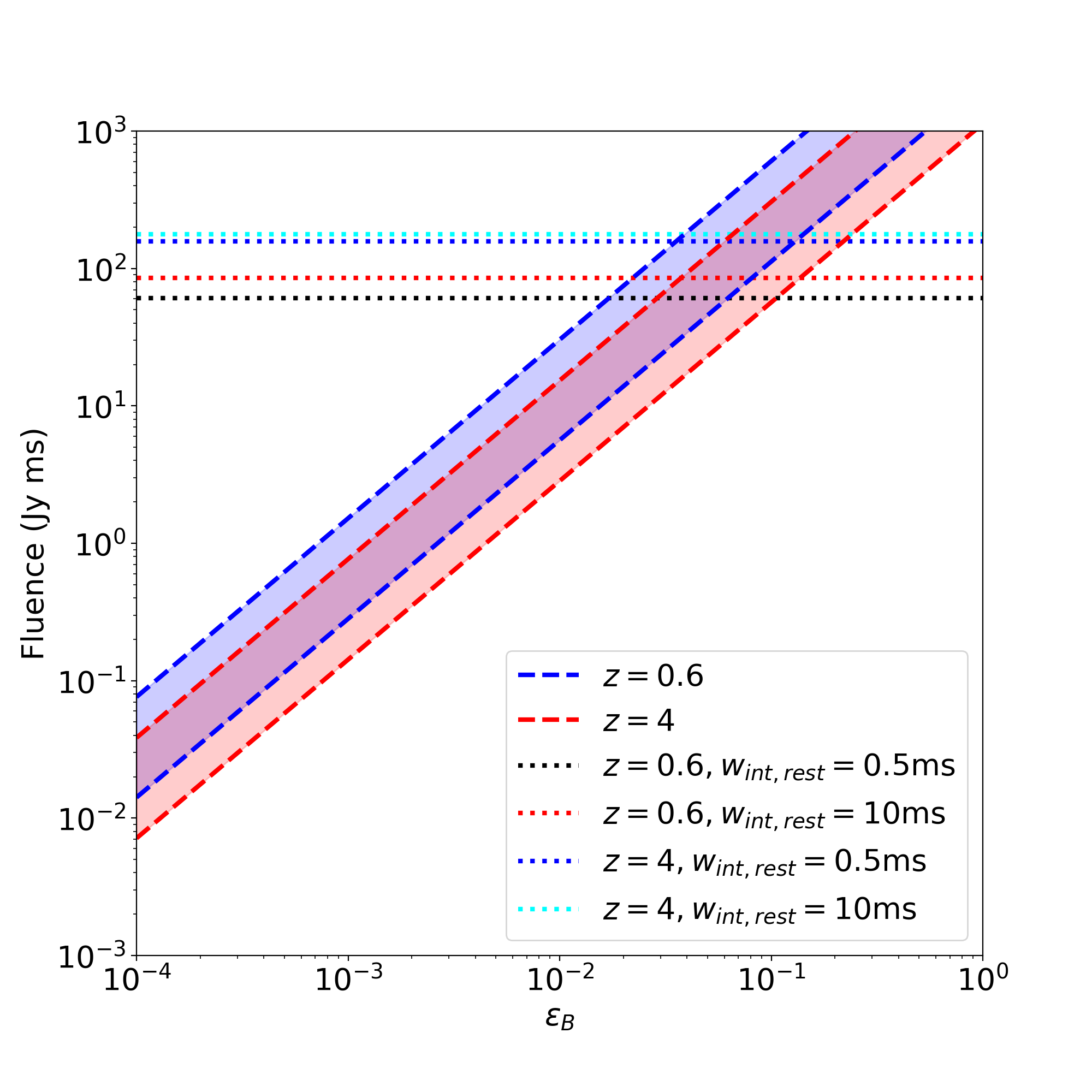}
\caption{The predicted fluence of a prompt signal produced by the interaction between the relativistic jet of GRB 210419A and the ISM at 185\,MHz as a function of the fraction of magnetic energy.
The shaded regions illustrate those predictions assuming the maximum and minimum redshift considered in this investigation, with the uncertainties resulting from the peak frequency of the prompt radio emission at the shock front (see Eq. \ref{eq:fpeak}) and the measured gamma-ray fluence (see Section~\ref{sec:Swift}), which has been corrected to a bolometric gamma-ray fluence (see Section~\ref{sec:model1}).
The horizontal dotted lines in different colors represent the fluence upper limits we obtained from the VCS observation of GRB 210419A
for different combinations of redshift and intrinsic pulse width.
} 
\label{model}
\end{figure}

With the assumed redshifts and intrinsic pulse widths, we illustrate how our fluence upper limits derived from our MWA observation of GRB 210419A can constrain the model predictions for the fraction of magnetic energy in the relativistic jet in Figure \ref{model}. For a redshift range of $0.6<z<4$ and an intrinsic pulse width of $0.5\,\text{ms}<w_{\text{int,rest}}<10\,\text{ms}$, we derived a $6\sigma$ fluence upper limit of 77--224\,Jy\,ms (see Section~\ref{sec:sensi}), resulting in a constraint on the fraction of magnetic energy in the relativistic jet launched by GRB 210419A $\epsilon_{\text{B}}\lesssim0.05$ and $\epsilon_{\text{B}}\lesssim0.1$ at the lowest and highest redshift, respectively.
These upper limits on $\epsilon_{\text{B}}$ are comparable to $\epsilon_{\text{B}}\lesssim[0.24\,\text{--}\,0.47]$ derived in \citet{Rowlinson19b} for long GRB 180706A. Note that our constraints on $\epsilon_\text{B}$ are only valid if the jet-ISM interaction is indeed active in the GRB under study.

As one of the key open questions in the GRB field, i.e. whether the relativistic jet is Poynting flux or baryon dominated, 
GRB jet magnetisation has been investigated extensively (e.g. \citealt{Lyutikov03}, \citealt{Begu15}, \citealt{Peer17}). \citet{Zhang09b} reported a lower limit of $\epsilon_{\text{B}}\gtrsim[0.94\,\text{--}\,0.95]$ at the photosphere radius based on the non-detection of a thermal component in gamma rays ($\sim50$\,keV) from GRB 080916C. Note that $\epsilon_{\text{B}}$ evolves with the radius from the central engine and may become much smaller at the deceleration radius where the relativistic ejecta collides into the ISM \citep{Kumar15}. A detailed simulation of spectra of GRB prompt emission using a hybrid relativistic outflow containing both fireball and Poynting-flux components finds $\epsilon_{\text{B}}\gtrsim0.5$ at a distance of $10^{15}$\,cm from the central engine (a possible prompt gamma-ray emission site covered by our MWA observation; \citealt{Gao15}).
Therefore, our constraint on the magnetisation of GRB jets potentially undermines the Poynting flux dominated scenario investigated in this simulation but at a low significance, particularly given our assumptions on the spectral index $\beta$, the GRB redshift and the pulse width.

\subsubsection{Radio emission during the X-ray flare}\label{sec:model2}

As the X-ray light curve of GRB 210419A displays flaring activity as shown in Figure~\ref{X-ray} (shaded region), we explore the GRB jet properties in the context of any radio emission associated with X-ray flaring in this section.

While X-ray flares are commonly observed following GRBs, their physical origin still remains unclear, with suggestions including internal dissipation (prompt-emission-like; \citealt{Falcone07, Chincarini10, Margutti10}) and external shock (afterglow-like; \citealt{Giannios06, Panaitescu06, Bernardini11}) mechanisms. There is a criterion to distinguish these two scenarios based on the flare variability and occurrence time ($\Delta t_\text{FWHM}$ the full width at half maximum of the pulse and $t_\text{pk}$ the time of the flare maximum; \citealt{Ioka05, Lazzati07}). If $\Delta t_\text{FWHM}/t_\text{pk}<1$, as is the case for GRB 210419A (see Figure~\ref{X-ray}), the flare is difficult to accommodate within the external shock model. Therefore, here we assume
an internal shock origin for the flare observed in the X-ray light curve of GRB 210419A (the same as for the prompt gamma-ray emission), which means the jet-ISM interaction model discussed in Section~\ref{sec:model1} may also apply to the X-ray flare \citep[as described by][]{Starling20}.

For this scenario, we use the X-ray fluence derived for the X-ray flare in Section~\ref{sec:Swift} to calculate the fluence of the predicted radio signal
using the \citet{Usov00} model equations given in Section~\ref{sec:model1}.
In order to convert the X-ray fluence measured in the $0.3-10$\,keV energy band to a bolometric gamma-ray fluence ($0.1-10^{4}$\,keV), we again used WebPIMMS and the power law spectral fit to the \textit{Swift}-XRT data provided by \citet[][see also Section~\ref{sec:Swift}]{Beardmore21} to derive a correction factor of $4.0^{+2.0}_{-0.8}$. Note that we used the photon index from the spectrum derived from the PC mode observation as the recorded data covers the duration of  
the X-ray flare we are investigating (see Figure~\ref{X-ray}).

When placing fluence limits on the associated radio emission, we do not consider signals of millisecond duration for this scenario. Given that the fluence from the X-ray flare is much lower than what is supplied by the prompt gamma-ray emission, it would provide a less stringent constraint on $\epsilon_{B}$ (see Eq.~\ref{eq:model}) than that calculated in Section~\ref{sec:model1}.
However, it is possible that the predicted radio pulse has a much longer duration, similar to that of the X-ray flare \citep{Starling20}. Any signals on such a long timescale would not be dispersion limited at reasonable GRB redshifts, and would have a flux density equal to the undispersed pulse (see eq. 16 in \citealt{Rowlinson19}).
We can therefore readily search for associated radio emission in an MWA image created over the same timescale as the X-ray flare duration.

\begin{figure}
\centering
\includegraphics[width=.5\textwidth]{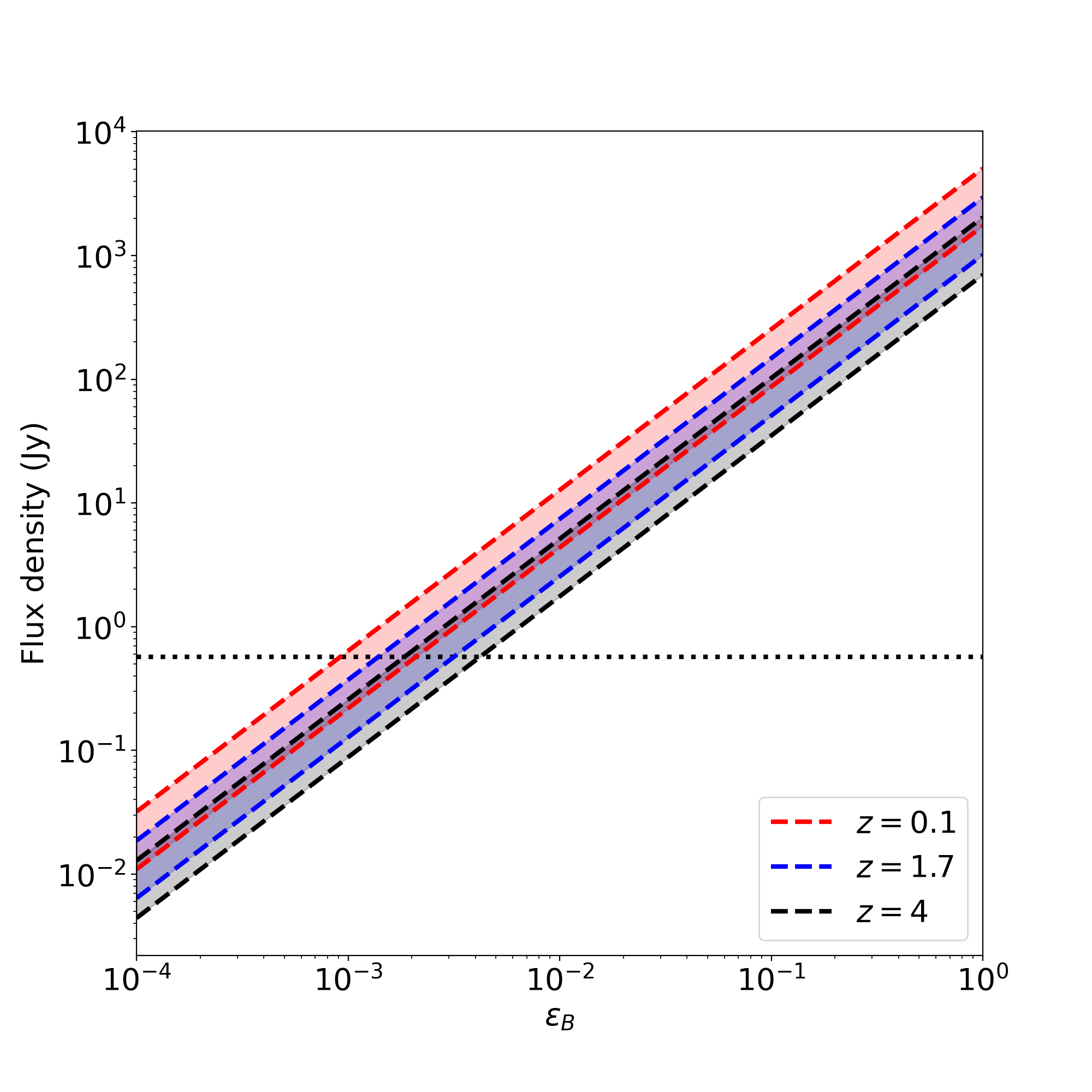}
\caption{The predicted flux density of the radio signal produced during the X-ray flare from GRB 210419A as a function of the fraction of magnetic energy. The shaded region in different colors represent the model predictions assuming the lowest, typical and highest long GRB redshift, with the uncertainties again resulting from the predicted peak frequency of the prompt radio emission at the shock front (see Eq.~\ref{eq:fpeak}) and the measured X-ray fluence (see Section~\ref{sec:Swift}), which has been corrected to a bolometric gamma-ray fluence (see Section~\ref{sec:model2}). The horizontal dotted line shows the $3\sigma$ flux density upper limit derived from the MWA image integrated over the duration of the X-ray flare.} 
\label{model_flare}
\end{figure}

In Figure~\ref{image} we show the region surrounding GRB 210419A made from an offline correlation of the VCS data with an integration time that covers the duration of the X-ray flare, assuming a
dispersion delay corresponding to a typical long GRB redshift of $z=1.7$ (see Section~\ref{sec:imaging}).
We compare the $3\sigma$ flux density upper limit derived from the MWA image in Figure~\ref{image} 
(which is similar to the upper limits from the images that assume redshifts of $z=0.1$ and $z=4$) 
to the predicted model emission associated with the X-ray flare 
for a range of redshifts 
in Figure~\ref{model_flare}. 
The MWA was on target and observing GRB 210419A before 
the X-ray flare, which occurred 335\,s post-burst 
so there is no lower limit on the redshift range we are able to constrain (unlike in Section~\ref{sec:model1}). 
We therefore plot the model predictions corresponding to the lowest ($z=0.1$), typical ($z=1.7$), and highest ($z=4$) observed long GRB redshifts. As can be seen from Figure~\ref{model_flare}, we are able to constrain the fraction of magnetic energy to $\epsilon_\text{B}\lesssim10^{-3}$, $\epsilon_\text{B}\lesssim2\times10^{-3}$ and $\epsilon_\text{B}\lesssim3\times10^{-3}$ during the flaring activity at the lowest, average, and highest redshifts. 
These constraints are more stringent than those
derived during the prompt gamma-ray emission phase in Section~\ref{sec:model1}. Note that our constraints on $\epsilon_\text{B}$ are only valid under the assumption that there is indeed radio emission during the X-ray flare of GRB 210419A.

\citet{Starling20} predicted that 44\% of X-ray flares detected by \textit{Swift}\textendash XRT should have had detectable low frequency radio emission by LOFAR assuming magnetically dominated GRB jets. Here, assuming a magnetic energy fraction of $\epsilon_\text{B}=10^{-2}$ comparable to the constraint shown in Figure~\ref{model_flare}, our MWA rapid-response observation should be able to detect the predicted radio emission from 30\% of X-ray flares.
Assuming a magnetically driven outflow at the base of the jet where it is launched (e.g. \citealt{Komissarov09,Tchekhovskoy10}), our non-detection of radio emission during the X-ray flare might imply the existence of magnetic energy dissipation in the GRB jet, which results in insufficient magnetic energy for radio emission during X-ray flares (e.g. \citealt{Kumar15}).

\subsection{Future prospects}

\subsubsection{Improvements to future VCS triggers}\label{sec:future}

We expect there to be much more sensitive observations with the full MWA in the future.
During our observation of GRB 210419A, several of the receivers were down due to beamformer faults on-site, which resulted in a 35\% sensitivity loss. With the full MWA operational, we could have reached a sensitivity of $\sim40$\,Jy\,ms for a 10\,ms wide pulse, comparable to the prediction in \citet{Rowlinson19}.
This would represent a factor of $\sim3$ in improvement in sensitivity compared to our results for GRB 210419A, which
would further constrain the fraction of magnetic energy in the relativistic jet of GRBs during the prompt gamma-ray phase (see Section~\ref{sec:model1}) to $\epsilon_{\text{B}}\lesssim[0.01\,\text{--}\,0.03]$ under the assumption that the jet-ISM interaction indeed operates in the GRB under study.

We expect to trigger the VCS on more \textit{Swift} GRBs in the future, with a particular focus on short GRBs
as associated prompt radio signals are more likely to escape their less dense surrounding environments \citep{Zhang14}. With a compact binary merger origin, short GRBs have additional channels to produce coherent radio emission such as the interactions of the neutron star magnetic fields just preceding the merger \citep{Lyutikov13}. Assuming a typical short GRB redshift of $z=0.7$ \citep{Gompertz20}, the MWA response time of 89\,s would allow us to capture the signals produced as early as $\sim13$\,s prior to the prompt gamma-ray emission. Based on the number of short GRBs detected by \textit{Swift} per year ($\sim9$; \citealt{Lien16}) and assuming 30\% sky coverage of the MWA, we would expect to trigger on $2\,\text{--}\,3$ short GRBs per year.

While the VCS data are most sensitive to prompt radio emission, they can be used to search for long timescale or persistent emission after offline correlation and imaging, as was done in Section~\ref{sec:model2}. However, since the observation of GRB 210419A was taken when the MWA was in the compact configuration, the resulting image has a low angular resolution ($\sim10$\,arcmin) and is limited by classical confusion (see Figure~\ref{image}; \citealt{Condon74, Franceschini82}), making the upper limit derived from the RMS noise less constraining than what we would expect from an observation taken in the extended configuration.
Assuming a typical flare duration of 247\,s (for the distribution of flare durations see \citealt{Yi16}) and the general relation that image noise scales with integration time as $\propto\Delta t_\text{int}^{-1/2}$, we expect a sensitivity of $\sim0.1\,\text{Jy}\,\text{beam}^{-1}$ on the 247\,s timescale in the extended configuration (a factor of 2 times better than our observation of GRB 210419A in the compact configuration) based on the upper limits derived on 30\,min timescales from previous MWA observations \citep{Anderson20, Tian22}. Note that our sensitivity estimation does not take into account sidelobe confusion.
In the future we expect to undertake VCS observations in the extended configuration, which will increase the sensitivity to any long timescale emission by an order of magnitude and thus improve our constraint on the coherent radio emission associated with X-ray flares.

\section{summary and conclusions}

In this paper, we have searched for prompt radio bursts associated with GRB 210419A in the frequency range of 170 to 200\,MHz using the rapid-response mode on the MWA, triggering VCS observations. This is the first time that the MWA VCS 
has been used in the rapid-response follow up of a GRB. 
The MWA rapid-response observing mode 
makes it possible to capture the early time emission, which would be missed by other low frequency radio telescopes with slower response times and/or all-sky instruments that necessarily have lower sensitivities (for a comparison of response times for different low-frequency telescopes see figure 10 in \citealt{Anderson20}).




As a result of this work, we come to the following main conclusions:

\begin{enumerate}

\item We have performed a single pulse search on the high time resolution data but found no prompt emission associated with GRB 210419A.
We derive a fluence upper limit of $77\text{--}224$\,Jy\,ms on prompt radio bursts associated with GRB 210419A, assuming a pulse width of $0.5\text{--}10$\,ms and a redshift of $0.6<z<4$. This allows us to test the jet-ISM interaction model assuming a spectral index of $\beta=1.6$ \citep{Usov00}.
The fluence limit results in the fraction of magnetic energy constraint of $\epsilon_{\text{B}}\lesssim[0.05\,\text{--}\,0.1]$ in the relativistic jet 
(see Figure~\ref{model}), disfavoring the Poynting flux dominated composition for the jet though at a low significance. 

\item We have also inspected the MWA images made via offline correlation of the VCS data for signals occurring during the X-ray flare of GRB 210419A assuming redshifts of $z=0.1$, 1.7, and 4 but found no emission at the GRB position (see Figure~\ref{image}), obtaining a $3\sigma$ flux density upper limit of $570\,\text{mJy}\,\text{beam}^{-1}$. This allows us to test the same jet-ISM interaction model, which also predicts radio emission during X-ray flares \citep{Starling20}.
The flux density limit results in a constraint on the magnetic energy fraction during the X-ray flare of $\epsilon_\text{B}\lesssim10^{-3}$ over a redshift range of $0.1<z<4$ (see Figure~\ref{model_flare}), suggesting magnetic energy dissipation in the GRB jet. 

\item Compared to previous MWA searches for prompt radio bursts using the standard correlator with a temporal resolution of only 0.5\,s \citep{Anderson20, Tian22}, our VCS observation of GRB 210419A with a temporal resolution of $100\,\upmu$s is
equally as sensitive to our best constrained burst GRB 190627A using image dedispersion techniques \citep{Tian22}, and demonstrates the potential for 
even more sensitive VCS observations in the future. 
\end{enumerate}

In conclusion, our non-detection of coherent radio emission associated with GRB 210419A seems to challenge the Poynting flux dominated scenario commonly assumed for GRB jets (\citealt{Usov94}; \citealt{Thompson94}; \citealt{Drenkhahn02}), which is a prerequisite for the radio emission mechanisms proposed by \citet{Usov00} and \citet{Starling20}.
However, there are some other possible reasons for our non-detection. Given the unknown redshift of GRB 210419A and observations of long GRBs at redshifts of $z>6$ \citep{Salvaterra15}, it may be too distant to have detectable radio emission. 
Given the X-ray absorption might not reflect the true density in the GRB environment (e.g. \citealt{Rahin19, Dalton20}), it is possible that GRB 210419A resides in a high density surrounding medium that prevents low-frequency emission from escaping.

In order to detect the predicted radio emission or fully explore the parameter space of the emission model, we need more MWA rapid-response VCS observations of GRBs, especially short GRBs with redshift measurements.

\section*{Acknowledgements}

J.T. would like to thank S. J. McSweeney for useful discussions regarding array
factor calculations, array efficiencies, and flux density calibration, and Chris Riseley for providing helpful feedback.

This scientific work makes use of the Murchison Radio-astronomy Observatory, operated by CSIRO. We acknowledge the Wajarri Yamatji people as the traditional owners of the Observatory site. Support for the operation of the MWA is provided by the Australian Government (NCRIS), under a contract to Curtin University administered by Astronomy Australia Limited.

This work made use of data supplied by the UK {\it Swift} Science Data Centre at the University of Leicester and the {\it Swift} satellite. {\it Swift}, launched in November 2004, is a NASA mission in partnership with the Italian Space Agency and the UK Space Agency. {\it Swift} is managed by NASA Goddard. Penn State University controls science and flight operations from the Mission Operations Center in University Park, Pennsylvania. Los Alamos National Laboratory provides gamma-ray imaging analysis. 

This work was supported by resources provided by the Pawsey Supercomputing Centre with funding from the Australian Government and the Government of Western Australia.

GEA is the recipient of an Australian Research Council Discovery Early Career Researcher Award (project number DE180100346), and JCAM-J is the recipient of an Australian Research Council Future Fellowship (project number FT140101082) funded by the Australian Government.

The following software and packages were used to support this work: {\sc mwa\_trigger} \citep{Hancock19}\footnote{https://github.com/MWATelescope/mwa\_trigger/},
{\sc comet} \citep{Swinbank14}, 
{\sc voevent-parse} \citep{Staley16},
{\sc Astropy} \citep{TheAstropyCollaboration2013,TheAstropyCollaboration2018}, 
{\sc numpy} \citep{vanderWalt_numpy_2011}, 
{\sc scipy} \citep{Jones_scipy_2001}, 
{\sc matplotlib} \citep{hunter07}, {\sc presto} \citep{Ransom01},   docker\footnote{\href{https://www.docker.com/}{https://www.docker.com/}}, 
singularity \citep{kurtzer_singularity_2017}, Ned Wright's Cosmology Calculator\footnote{http:/www.astro.ucla.edu/$\sim$wright/CosmoCalc.html}~\citep{Wright06}.
This research has made use of NASA's Astrophysics Data System. 
This research has made use of SAOImage DS9, developed by Smithsonian Astrophysical Observatory.
This research has made use of the VizieR catalogue access tool \citep{ochsenbein00} and the SIMBAD database \citep{wenger00}, operated at CDS, Strasbourg, France.

\section*{Data availability}

The MWA data used are available in the MWA Long Term Archive at \url{http://ws.mwatelescope.org/metadata/find} under project code D0009 and PI J Tian. The time series of Stokes parameters formed at the position of GRB 210419A have been uploaded to Zenodo (\mbox{\url{https://doi.org/10.5281/zenodo.5821351}}). The Swift data used are available from the UK Swift Science Data Centre at the University of Leicester at \mbox{\url{https://www.swift.ac.uk/index.php}} under GRB 210419A.




\bibliographystyle{mnras}
\bibliography{bib} 



\appendix

\section{candidates of dispersed pulse search}\label{appendix:candi}

\begin{figure*}
\centering
\subfigure{
  \label{p}
  \includegraphics[width=.47\linewidth]{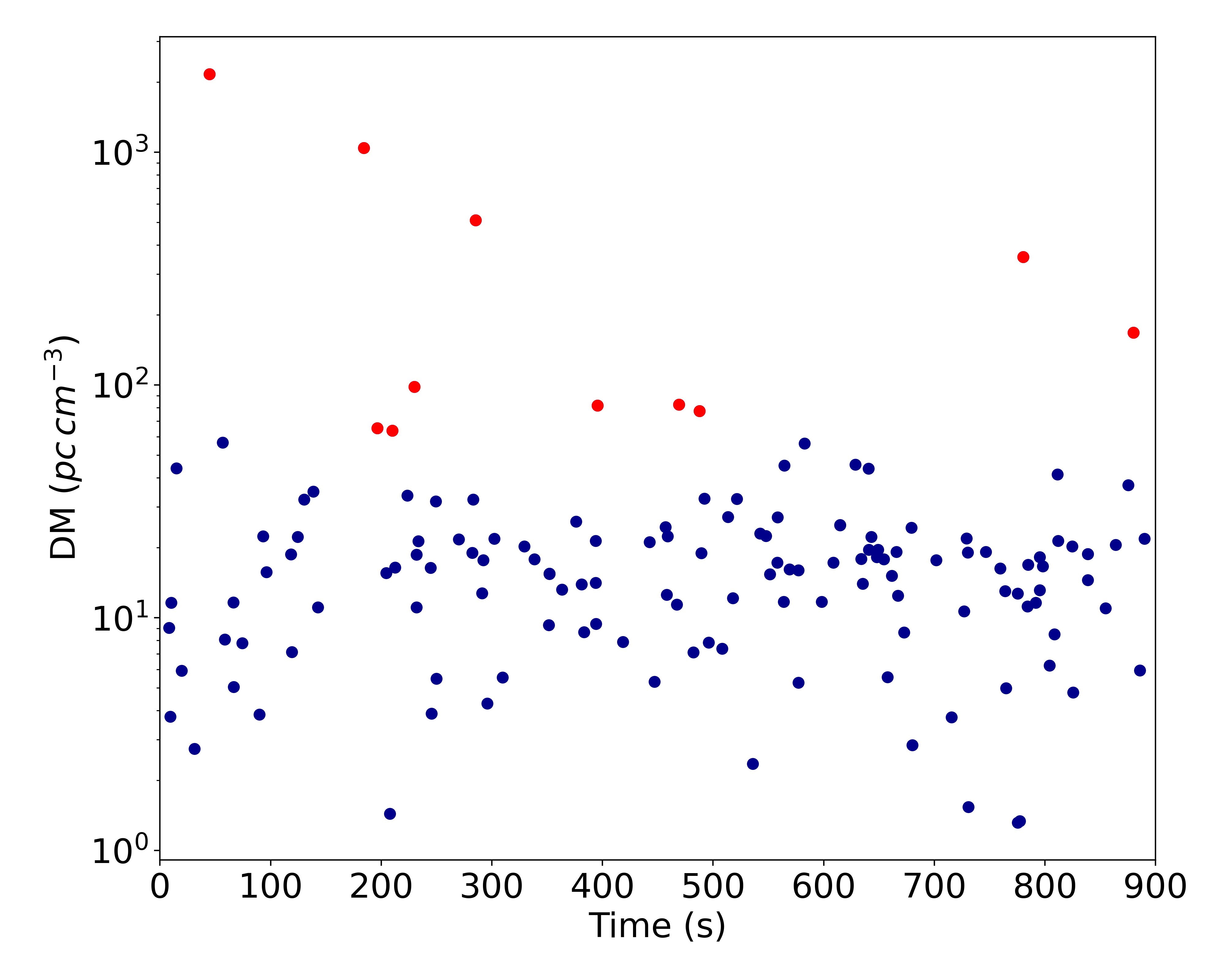}}
\qquad  
\subfigure{
  \label{n}
  \includegraphics[width=.47\linewidth]{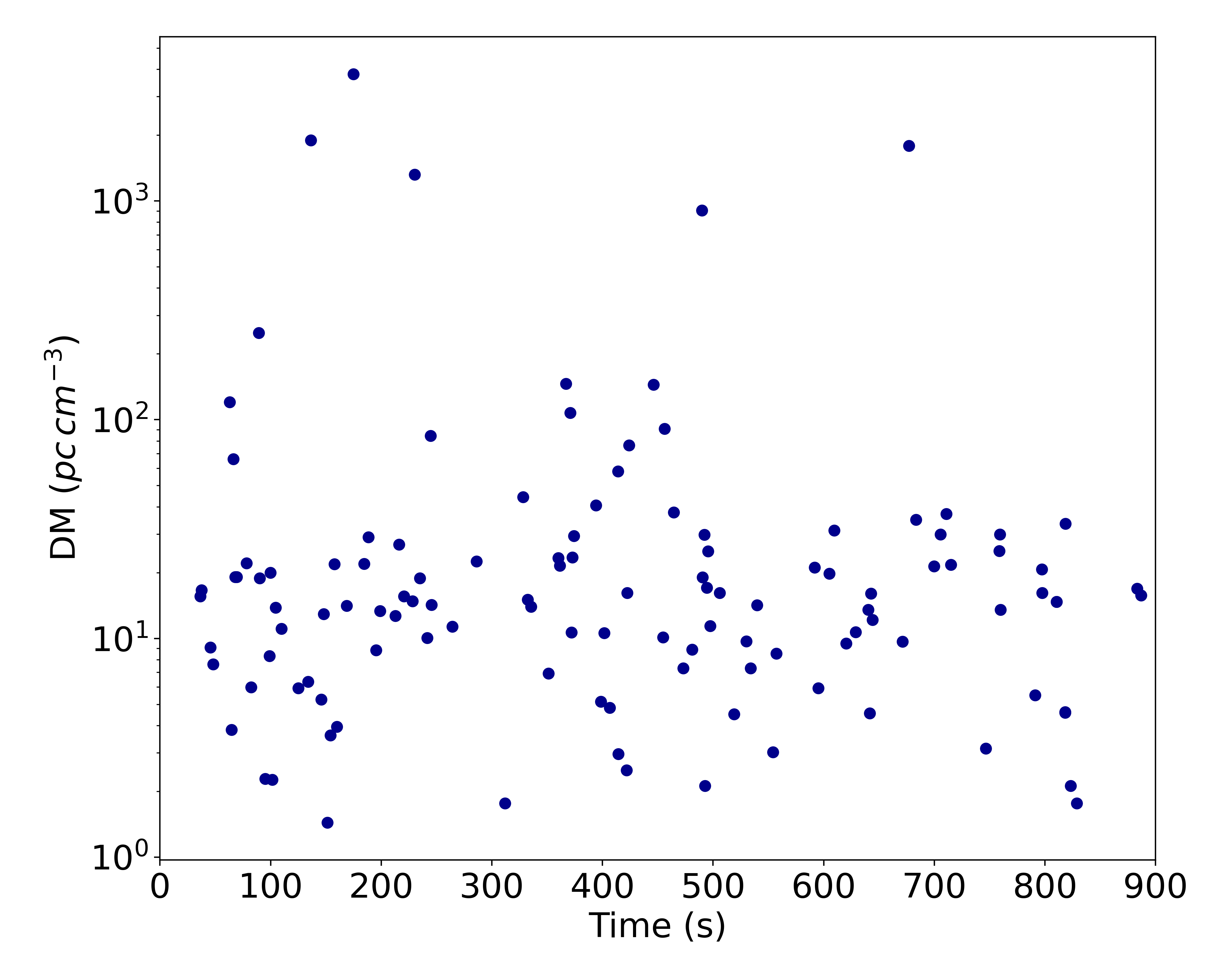}}
\caption{Single pulse candidates with SNR above $6\sigma$ (blue circles) produced by positive (left) and negative (right) DM trials, respectively. The 11 candidates with $\text{DM}>62\,\text{pc}\,\text{cm}^{-3}$ are marked with red colors.
}
\label{cand}
\end{figure*}

In Figure~\ref{cand}, we provide the candidates output by {\sc presto} with SNR above $6\sigma$ from our dispersed pulse search on the rapid-response MWA VCS observation of GRB 210419A. The candidates found from another set of time series created on the same dataset using (unphysical) negative DM trials are also included for comparison (see Section~\ref{sec:results1}).

\begin{figure*}
\centering
\subfigure{
  \label{hist_l}
  \includegraphics[width=.47\linewidth]{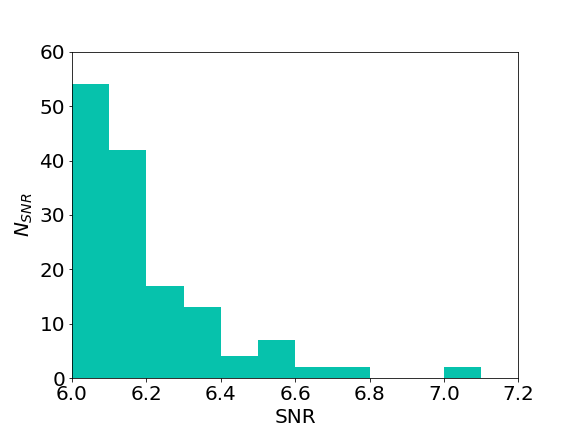}}
\qquad  
\subfigure{
  \label{hist_r}
  \includegraphics[width=.47\linewidth]{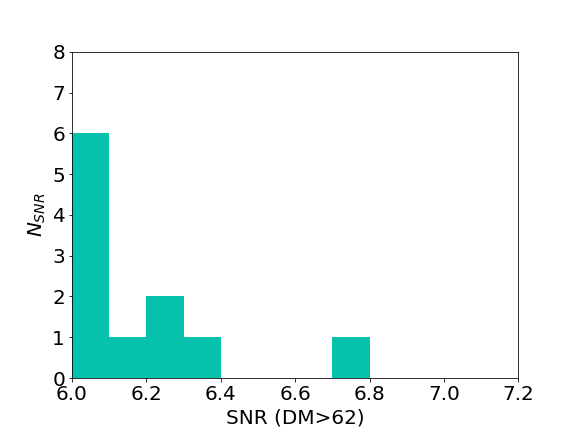}}
\caption{Distribution of SNRs of all candidates above $6\sigma$ (left) and those with $\text{DM}>62\,\text{pc}\,\text{cm}^{-3}$ (right), respectively.
}
\label{hist}
\end{figure*}

In Figure~\ref{hist}, we present the distribution of SNRs of all candidates shown in the left panel of Figure~\ref{cand}. We also plot a histogram for those candidates with $\text{DM}>62\,\text{pc}\,\text{cm}^{-3}$ (red points in Figure~\ref{cand}), which are more likely to originate from cosmological distances (see Section~\ref{sec:search}).

\bsp	
\label{lastpage}

\end{document}